# Challenges in Comparing Code Maintainability across Different Programming Languages

Christophe Ponsard, Gustavo Ospina and Denis Darquennes

*CETIC Research Centre, Gosselies, Belgium*

**Abstract**

Comparing the quality of software written in different computer languages is required in a variety of scenarios, e.g. multi-language projects or application selection process among candidates in different languages. We focus on the challenges related to comparing the maintainability quality typically through a maintainability index or technical debt approaches. We identify and discuss how to manage a number of challenges to produce comparable maintainability assessments across languages related to the programming paradigm (purely procedural vs OO vs multi-paradigm), the coverage of key quality dimensions, and the use of generic metrics vs more languages specific rules. Our work is based on a set of code analysis carried out in Wallonia over the past 15 years.

## 1. Context and Objective

Software maintainability is the quality of software enabling it to be easily modified without breaking existing functionality [1]. Software maintenance is considered as a costly phase in the software life cycle with estimates up to 70% of global project time and resources for this activity [2]. The need for maintenance is also stressed by Lehman's second law of software evolution.

Assessing the maintainability is well supported by standards such as SQuaRE defining different attributes: modularity, reusability, analysability, modifiability and testability [3] and evaluation methods based on static analysis of code. These cover a wider range of qualities such as security and reliability but primary focus on maintainability based on the evaluation of specific quality metrics in connection with an underlying quality model (i.e. measuring quality) or on the notion of technical debt based on a set of coding rules resulting in various types of defects or code smells (i.e. measuring non-quality) [4].

The focus of our work is on specific difficulties arising when multiple programming languages are involved. In contrast with situations where the same language is monitored over time on a project or across a project portfolio, the comparability is more challenging because impacted by the language and the methodology used, e.g. the formulation of metrics like code duplication in terms of repeated lines of code/tokens vs the language verbosity, the estimate of the production effort for technical debt ratio, different rule sets for code smells across languages or tools.

We encountered a variety of such scenarios while helping Walloon companies with development practices. Some examples used to illustrate our work are:

- prioritising maintenance activities within a project portfolio, e.g. in public sector (Walloon DTIC).
- application selection among functionally similar candidates, e.g. to migrate a sector using a variety of mature solutions (>20 years old) ranging from a COBOL-like 4GL to C# and Java.
- monitoring the maintainability during software modernisation, e.g. in a safety critical application.
- comparing the development practices in Wallonia through code audits over the past 15 years [5].

Our objective is to identify interesting challenges and to share our insights about how to manage them practically with anonymised illustrations on our scenarios and connections with related work.

## 2. Outline

Our work reviews representative approaches for maintainability estimation and highlights challenges and some practical solutions relating to our comparison scenarios. We rely on three main categories[6].

**The Maintainability Index (MI)** and its variants rely on an empirically aggregation of global metrics. A typical aggregated indicator still used by Visual studio is as follows and combines averages of the Halstead volume (aVE), cyclomatic complexity (aCC) and LOC per module (aLOC). Optionally documentation ratio is also used. The MI raised many issues about the lack of transparency, the arbitrary weights and missing quality dimensions (e.g. duplication)[7] in line with our experience, e.g. we experienced inconclusive evaluation due to cancellation effects between different terms.

$$MI_VS = max(0, 100 \times \frac{171 - 5.2 \times ln(aHV) - 0.23 \times aCC - 16.2 \times ln(aLOC)}{171})$$

**The Software Improvement Group (SIG) model** maps system characteristics from the ISO9126 (now ISO25010 [3]) on source code properties, i.e. analysability, changeability stability and testability [7]. These are evaluated by language-independent metrics easy to apply in our scenarios. Those metrics partially overlap with MI metrics and cover volume, complexity, duplication, unit size and unit testing. Some discussed limitations relate to the Mc Cabe complexity metric, the possible breakdown of metrics on specific part of the architecture (e.g. business logic), coupling metrics for better assessing the architecture and criticisms about how to estimate the repair effort based on empirical factors.

**The Software QuALity Enhancement model (SQALE)** relies on the ISO 25010 Software Quality standard through several code metrics that are attached to its taxonomy. from main characteristics like maintainability to sub-characteristics like understandability/readability to be finally assessed by code level rules which can be related to code metrics (like complexity or method size associate with a threshold) or more specific rules like naming conventions or avoiding too many parameters in a method. Based on a given set of *rules* which is language dependant, the (lack of) maintainability is assessed using the technical debt ratio (TDR) based on the assumption all defects are corrected with the formula:

$$TDR = \frac{\sum_{r \in rules} \text{effortToFix}(violation(r))}{\text{ProductionEffort}}$$

The ratio can then be translated to a maintainability scale typically from A to E depending on the ratio: A=[0,5%], B=]5-10%], C=]10%,20%], D=]20%,50%], E=]50%,100%]. We discuss here issues related to the set of rules used across languages that might not be comparable by potentially generating different levels of technical debt, on the estimation of the production effort which can be touchy to assess, and on the status of code duplication in some tools.

# Challenges in Comparing Code Maintainability across Different Programming Languages

Christophe Ponsard, Gustavo Ospina, Denis Darquennes - CETIC - christophe.ponsard@cetic.be

## Context
- Software maintainability : quality of software about ease of modification without breaking existing functionality
- The need for maintenance is stressed by Lehman's second law of software evolution
- Software maintenance: costly with estimates up to 70% of global project time and resources [1]

### Maintainability Assessment:
- SQuaRE ➔ quality model: maintainability and beyond
- Approaches: measuring
  - <u>quality</u> using specific metrics,
    e.g. documentation, complexity, coupling,…
  - <u>non-quality</u>: technical debt
- Measured through Static Code Analysis tools : e.g. SonarQube, CAST

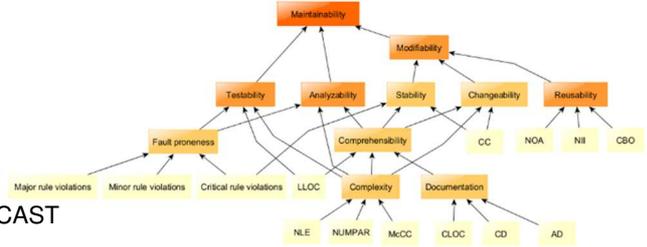

### Research question: what if multiple programming languages are involved ➔ some scenarios
- maintaining a project mixing multiple languages
- prioritizing maintenance in the scope of an heterogeneous project portfolio
- selecting best option among functionally equivalent candidates in different languages
- comparing development practices in a region/sector over a long time and across technologies

## Existing approaches (based on [2]) and limitations

### Maintainability Index (Oman and Hagemeister, 1991) – composite index
- aggregating Halstead volume (aVE), cyclomatic complexity (aCC) and LOC per module (aLOC)
- e.g. Visual Studio variant (still used) $MI_{VS} = max(0, 100 \times \frac{171 - 5.2 \times ln(aHV) - 0.23 \times aCC - 16.2 \times ln(aLOC)}{171})$
- <u>many issues [3]</u>: lack of transparency, arbitrary weights, missing qualities (e.g. duplication), …

### Software Improvement Group (SIG) model
- maps system characteristics from ISO9126/SQuaRE: analysability, changeability stability and testability
- based on language-independent metrics e.g. volume, complexity, duplication, unit size and unit testing
- <u>limitations</u>: McCabe complexity, missing coupling, too empirical repair effort

### Software QuALity Enhancement model (SQALE)
- 3 level taxonomy from ISO9126/SQuaRE, only level 3 is language dependent
- aggregated in a technical debt (cost to correct) ratio $TDR = \frac{\sum_{r \in rules} effortToFix(violation(r))}{ProductionEffort}$
- A to E scale based on TDR: A=[0,5%], B=]5-10%], C=]10%,20%], D=]20%,50%]
- <u>limitations</u>: reliability of estimation of production effort,
  completeness/granularity of rule set, status of code duplication

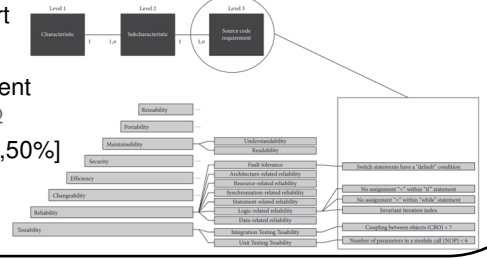

## Our practical approach

### Key principles:
- Use relative indicators, possibly take volume into account separately
- Rely on metrics (e.g. comment ratio) and rules (part of TDR)
- Make sure of comparability across the considered languages
  - basic indicators are fine: McCabe complexity, comment ratio,…
  - keep similar rules at intersection of rule sets
  - avoid pure OO metrics if mixing with procedural (e.g. COBOL)
- Use a consistent and well-defined internal aggregation process
  - define acceptable bounds and mapping to reference scale like 0..100
  - use simple explainable weights (not like MI)
  - count each attribute ONCE either through indicator or tech debt
- For production effort: use the same estimation method
- Compare indicators separately and together, consider sensitivity analysis
- Also compare with outcome of other approaches (MI, SQI, SQALE)

### Example: for functional selection scenario

| Indicator | Mapping on 0..100 | Weight |
|---|---|---|
| Comment ratio | 15%..40% (+decreasing after) | 15% |
| Duplication ratio | 15%..5% (inversed scale) | 15% |
| Tech debt ratio<br>- complexity<br>- unit size<br>- coding rules | <br>10..20<br>0..MAX (verbosity!)<br>naming, hiding,… | 45% (= 3x15%) |
| Volumetry | Min.. 1.5 x Min | 25% |

## Current status and next steps
Approach successful for comparison and explainability on a set of languages from legacy 4GL to current mainstream languages
Revealed some caveats in previous methods: e.g. tooling bias, duplication estimation
Planning to generalise on our database [4] and extend comparison scope (e.g. architecture indicators, testing)

## Key References